  \documentclass[final,3p,times,twocolumn]{elsarticle}

\usepackage{amssymb}
\usepackage{amsmath,amssymb,exscale}
\usepackage{graphicx}
\usepackage{epsfig}
\usepackage{multicol}
\usepackage{color}
\usepackage{hyperref}
\hypersetup{colorlinks,bookmarksopen,bookmarksnumbered,citecolor=rossos,
linkcolor=black,pdfstartview=FitH,urlcolor=blus}
\usepackage{slashed}

\definecolor{blus}{cmyk}{1,1,0,0.6}
\definecolor{verdes}{cmyk}{0.99,0,0.59,0.82}
\definecolor{rossos}{cmyk}{0,1,1,0.55}
\definecolor{greeny}{cmyk}{0.99,0,0.59,0.98}

\def\be{\begin{equation}}
\def\ee{\end{equation}}
\def\bea{\begin{eqnarray}}
\def\eea{\end{eqnarray}}

\definecolor{red}{rgb}{1,0,0}

\journal{the arXiv}

\begin{document}

\begin{frontmatter}



\title{Higgs Inflation at NNLO after the Boson Discovery}


\author{Alberto Salvio}

\address{Departamento de F\'isica Te\'orica, Universidad Aut\'onoma de Madrid\\ and Instituto de F\'isica Te\'orica IFT-UAM/CSIC,  Madrid, Spain. \\ {\it {\small Report numbers: FTUAM-13-22, IFT-UAM/CSIC-13-089}} }

\begin{abstract}
We obtain the bound on the Higgs and top masses to have Higgs inflation (where the Higgs field is non-minimally coupled to gravity) at full next-to-next-to-leading order (NNLO). Comparing the result obtained with the experimental values of the relevant parameters we find some tension, which we quantify. Higgs inflation, however, is not excluded at the moment as the measured values of the Higgs and top masses are close enough to the bound once experimental and theoretical uncertainties are taken into account.
\end{abstract}


\end{frontmatter}


\section{Introduction}

The discovery of the Higgs boson \cite{Higgsexp1, Higgsexp2} has allowed to fix the last Standard Model (SM) parameter, the Higgs mass. Making the strong but certainly economical assumption that the SM (appropriately extended to accommodate neutrino masses and dark matter) remains valid up to the Planck scale, it is now possible to obtain precise predictions in this vast energy range.

Ref. \cite{Bezrukov:2007ep} argued that even the inflationary period of the Universe can be explained within the SM  and the Higgs field and the inflaton can be identified if the 
term 
\begin{equation}
\sqrt{-g}\xi H^{\dagger} H R, \label{nonminimal}
\end{equation}
with $\xi\gg 1$, is added to the Einstein-Hilbert plus  SM Lagrangian $\mathcal{L}_{E-H}+\mathcal{L}_{SM}$, so that  the total Lagrangian is 
\be \mathcal{L}_{total}= \mathcal{L}_{E-H}+\mathcal{L}_{SM} +\sqrt{-g}\xi H^{\dagger} H R. \label{Jordan-Lagrangian}\ee
Here $R$ is the Ricci scalar, $H$ is the Higgs doublet and $g$ is the determinant of the metric $g_{\mu \nu}$.

An inflaton with a non-minimal coupling of the form given in (\ref{nonminimal}), and in particular Higgs inflation, is perfectly consistent with recent Planck results \cite{Ade:2013uln}, which favor a simple single field inflation.

All this reinforces the interest in the possibility of Higgs inflation.

The non-minimal coupling in (\ref{nonminimal}) can be eliminated by a redefinition of $g_{\mu \nu}$ (going to the so called Einstein frame), which leads to a non-polynomial Lagrangian  for $H$.  This redefinition shows that  two regimes are present in the theory \cite{Bezrukov:2008ej}: the small field one $|H|\ll M_P/\xi$, where the canonical SM is a good description, and the large field limit $|H|\gg  M_P/\xi$, in which the physical Higgs mode decouples. Therefore, the latter limit corresponds to the chiral electroweak (EW) theory \cite{Longhitano:1980iz}.

As we will review in section \ref{Classical analysis}, at the classical level this is a viable model of inflation if the non-minimal coupling $\xi$ is chosen to match cosmic microwave background (CMB) observations. Quantum corrections  may, however, render inflation impossible depending on the input parameters at the EW scale, in particular the Higgs and top pole masses $M_h$ and $M_t$: if $M_h$ is too small (or $M_t$ is too large) the slope of the Higgs effective potential at large field values becomes negative preventing the field configuration to roll towards the EW vacuum.

In this paper we improve on previous determinations \cite{Bezrukov:2008ej} of the lower bound on the Higgs mass (or equivalently the upper bound on the top mass) to have Higgs inflation by using the following ingredients: (1)  two loop effective potential in the inflationary regime  including the effect of $\xi$ and the leading SM couplings: the top Yukawa $y_t$, the strong gauge coupling $g_3$, the EW gauge couplings $g_2$ and $g_1$ and the quartic Higgs coupling $\lambda$; (2) three loop SM renormalization group equations (RGE) from the EW scale up to $M_P/\xi$  for $y_t$,  $g_3$,  $g_2$, $g_1$ and $\lambda$ including the effects of all these couplings; (3)  two loop RGE for the same SM couplings and one loop RGE for $\xi$ in the chiral EW theory; (4)  recent precise determinations of these SM couplings at the top mass provided in \cite{Buttazzo:2013uya}, which are used as initial conditions for the RGE \footnote{See, however, Ref. \cite{Antipin:2013sga} for a related possible issue if conformal invariance is required}. 

A detailed description of these ingredients is provided in section  \ref{Quantum corrections}.
In section \ref{numerical analysis} we present our numerical results, including the determination of $\xi$ and the lower bound on $M_h$ (or $M_t$).
Finally in section \ref{conclusions} we conclude.

\section{Classical analysis} \label{Classical analysis}
Let us briefly review the model of \cite{Bezrukov:2007ep} at the classical level. The part of the action in (\ref{Jordan-Lagrangian}) that depends on the metric and the Higgs field  {\it only} is 
\begin{equation} S_{gH} = \int d^4x\sqrt{-g}\left[\left(\frac{M_P^2 }{2}+\xi H^{\dagger} H\right)R+|\partial H|^2-V\right], \nonumber \end{equation}
where $M_P\simeq 2.435 \times 10^{18}$ GeV is the reduced Planck mass,  $V=\lambda (H^{\dagger} H-v^2/2)^2$ is the classical Higgs potential, and $v$ is the EW Higgs vacuum expectation value. 

The non-minimal coupling (\ref{nonminimal}) can be eliminated through the {\it conformal} transformation
\begin{equation} g_{\mu \nu}\rightarrow \hat{g}_{\mu \nu}\equiv  \Omega^2  g_{\mu \nu}, \quad \Omega^2= 1+\frac{2\xi H^{\dagger} H}{M_P^2}. \label{transformation}\end{equation}
The original frame, where the Lagrangian has the form in (\ref{Jordan-Lagrangian}), is called the Jordan frame, while the one where gravity is canonically normalized (obtained with the transformation above) is called the Einstein frame.
In the unitary gauge, where the only scalar field is the radial mode $\phi \equiv \sqrt{2H^\dagger H}$,  we have (after the conformal transformation)
\begin{equation} S_{gH} = \int d^4x\sqrt{-\hat{g}}\left[\frac{M_P^2 }{2}\hat{R}+K \frac{(\partial \phi)^2}{2}-\frac{V}{\Omega^4}\right], \end{equation}
where $K\equiv (\Omega^2+6\xi^2\phi^2/M_P^2)/\Omega^4$.
The non-canonical Higgs kinetic term can be made canonical through the field redefinition $\phi=\phi(\chi)$ defined by
\begin{equation} \frac{d\chi}{d\phi}= \sqrt{\frac{\Omega^2+6\xi^2\phi^2/M_P^2}{\Omega^4}}.\label{chi}\end{equation}
Thus, $\chi$ feels a potential 
\begin{equation} U\equiv \frac{V}{\Omega^4}=\frac{\lambda(\phi(\chi)^2-v^2)^2}{4(1+\xi\phi(\chi)^2/M_P^2)^2}\label{U} .\end{equation}
From (\ref{chi}) and (\ref{U}) it follows \cite{Bezrukov:2007ep} that $U$ is exponentially flat when $\chi \gg M_P$, which is a key property to have inflation. Indeed, for such high field values the parameters
\bea \epsilon \equiv\frac{M_P^2}{2} \left(\frac{1}{U}\frac{dU}{d\chi}\right)^2, \quad \eta \equiv \frac{M_P^2}{U} \frac{d^2U}{d\chi^2},\nonumber \\  \zeta^2\equiv \frac{M_P^4}{U^2}\frac{d^3U}{d\chi^3} \frac{dU}{d\chi} \label{epsilon-def}\eea
are guaranteed to be small. Therefore, the region in field configurations $\chi > M_P$ (or equivalently \cite{Bezrukov:2007ep} $\phi> M_P/\sqrt{\xi}$) corresponds to inflation.

All the parameters of the model can be fixed through experiments and observations, including $\xi$ \cite{Bezrukov:2007ep, Bezrukov:2008ut}, so that Higgs inflation is highly predictive and as such falsifiable. $\xi$ can be fixed by requiring that the WMAP normalization of \cite{Hinshaw:2012aka},
\begin{equation}\frac{U}{ \epsilon}=24\pi^2\Delta^2_R M_P^4\simeq (0.02746 M_P)^4, \label{normalization} \end{equation}
is reproduced for a field value $\phi=\phi_{WMAP}$ corresponding to an appropriate number of e-foldings \cite{Bezrukov:2008ut}:
\begin{equation}N=\int_{\phi_{\rm end}}^{\phi_{WMAP}}\frac{U}{M_P^2}\left(\frac{dU}{d\phi}\right)^{-1}\left(\frac{d\chi}{d\phi}\right)^2d\phi\simeq 59, \label{e-folds}\end{equation}
where $\phi_{\rm end}$ is the field value at the end of inflation, 
\be \epsilon(\phi_{\rm end}) \simeq 1. \label{inflation-end}\ee
 This procedure leads to $\xi \simeq 4.7\times10^4 \sqrt{\lambda}$,  which is why $\xi$ has to be much larger than one.
 
 We can also extract the spectral index $n_s$,  the tensor-to-scalar ratio $r$ and the running spectral index $dn_s/d\ln k$:
\bea n_s&=&1-6\epsilon +2\eta, \nonumber \\ r&=&16\epsilon \\ \frac{dn_s}{d\ln k}&=&16 \epsilon \eta- 24\epsilon^2  -2\zeta^2, . \nonumber  \eea
These parameters are of interest as they are constrained by observations \cite{Ade:2013uln}.

\section{Quantum corrections} \label{Quantum corrections}

We now turn to the quantum corrections. 
We will use perturbation theory to compute them. It is important to keep in mind that perturbative unitarity\footnote{This unitarity problem can be solved by adding an extra real scalar field \cite{Giudice:2010ka}. The extension of the present analysis to include such scalar is beyond the scope of this paper.} is violated above some high energy scale \cite{crit,Burgess:2010zq}. Once the background fields are taken into account, however, one can show \cite{Bezrukov:2010jz} that such energy is parametrically higher than all relevant scales during the history of the Universe. Nevertheless some additional assumptions on the underlying ultraviolet  complete theory are necessary (see \cite{Burgess:2010zq,Bezrukov:2010jz, Bezrukov:2012sa}). 

There are two options for the quantization of the classical theory defined before: one can either first perform the transformation in (\ref{transformation}) and then quantize \cite{Bezrukov:2007ep} (prescription I) or first quantize and then perform the conformal transformation (prescription II) \cite{DeSimone:2008ei}. The two options lead to  different theories as they have different predictions \cite{Bezrukov:2008ej}.  We choose the first possibility because Ref. \cite{Bezrukov:2008ej} found it to be the one leading to the weaker bound on $M_t$ and such bound, as we will see, is already giving some tension with the experiments. We will make some more comments on prescription II at the end of section \ref{numerical analysis}, where we will check that it is indeed leading to a stronger bound even at full NNLO.

The procedure to compute quantum corrections has been introduced in \cite{Bezrukov:2008ej}: we briefly summarize it in the following subsections giving both the order of approximation reached in \cite{Bezrukov:2008ej} and our improvements.

\subsection{Effective potential}

The first element that we need is the (quantum) effective potential for $\chi$, which is expanded in loops as $$U_{\rm eff}=U+U_1+U_2+...\, . $$ Here $U$ is the classical contribution in Eq. (\ref{U}) and $U_1$, $U_2$, ... are the one loop, two loop, ... contributions respectively. An observation that leads to useful simplifications  is that we only need $U_{\rm eff}$ in the inflationary regime. Also, further simplifications can be achieved with a judicious gauge choice; we choose the Landau gauge. 

 $U_{\rm eff}$ depends mainly on the top, W,  Z, physical Higgs and (would-be) Goldstone squared masses in the classical background $\phi$ \cite{Coleman:1973jx}, which we call $t$, $w$, $z$, $h$ and $g$ respectively. We have\footnote{Note that we find some differences in the expressions of $h$ and $g$ with respect to those in \cite{Allison:2013uaa}.}
\bea   t\equiv \frac{y_t^2\phi^2}{2\Omega^2} , \quad w\equiv \frac{g_2^2\phi^2}{4\Omega^2}, \quad z\equiv \frac{(g_2^2+3g_1^2/5)\phi^2}{4\Omega^2}\nonumber  \\  h \equiv \frac{3\lambda \phi^2(1-\xi \phi^2/M_P^2)}{\Omega^4(\Omega^2+6\xi^2 \phi^2/M_P^2)} ,\quad  g\equiv \frac{\lambda\phi^2}{\Omega^4},\nonumber \eea
where we neglected $v$, whose contribution is amply negligible in the inflationary regime. Note that $h$ becomes negative for $\phi>M_P/\sqrt{\xi}$,  which also follows from the fact that $U$ is asymptotically flat; this problem can be ignored because  $h$ is negligible: it is suppressed (compared to $t$, $w$ and $z$) by an extra power of $\xi \phi^2/M_P^2$ in the inflationary regime and by an extra power of $\xi$; also, $g$ has $\Omega^4$ rather $\Omega^2$ in the denominator, which implies that it is suppressed  in the deep inflationary regime. Thus, in practice, we  obtain (as in \cite{Bezrukov:2008ej}) that the most relevant squared masses are $t$, $w$ and $z$.

So the one loop part is well approximated by\footnote{In the one loop part  $U_1$we kept the contribution of $g$ because it may modify the effective potential at the end of inflation if $\lambda$ is not too small. The input parameters at the EW scale, however, correspond to small values of $\lambda$ during the whole period of inflation, such that  this contribution will be negligible.} (in the modified minimal subtraction ($\overline{\rm MS}$) scheme) 
\bea \hspace{-0.5cm}  U_1 =  \frac{1}{(4\pi)^2}\left[\frac{3}{2}w^2\left(\ln\frac{w}{\bar{\mu}^2}-\frac{5}{6}\right)+\frac{3}{4}z^2\left(\ln\frac{z}{\bar{\mu}^2}-\frac{5}{6}\right)\right.\nonumber \\ && \hspace{-7cm} \left.-3t^2\left(\ln\frac{t}{\bar{\mu}^2}-\frac{3}{2}\right)+\frac{3}{4}g^2\left(\ln \frac{g}{\bar{\mu}^2}-\frac{3}{2}\right)\right], \nonumber  \eea
where $\bar{\mu}$ is the $\overline{\rm MS}$ renormalization scale.
 Ref. \cite{Bezrukov:2008ej} obtained the two loop effective potential in the inflationary regime by taking the $\overline{\rm MS}$ SM two loop effective potential in the Landau gauge, presented in \cite{Ford:1992pn}, dropping all diagrams involving the physical Higgs field (which decouples during inflation) and setting $g=0$. We do the same here. Therefore, our expression for $U_{\rm eff}$ in practice is  the one considered in \cite{Bezrukov:2008ej}.

\subsection{Renormalization group equations}
We RG-improve $U_{\rm eff}$ by using the running $\overline{\rm MS}$ couplings $\lambda(\bar{\mu})$, $y_t(\bar{\mu})$, $g_3(\bar{\mu})$, $g_2(\bar{\mu})$, $g_1(\bar{\mu})$ and $\xi(\bar{\mu})$. In order to keep the logarithms in the effective potential small we choose 
\be \bar{\mu}=\frac{\phi}{\Omega_t}\equiv \frac{\phi}{\sqrt{1+\xi_t\phi^2/M_P^2}},\label{prescriptionI}\ee
 where $\xi_t$ is $\xi$ evaluated at some reference energy (see below for its actual value in the numerical studies).

We compute the running of the $\overline{\rm MS}$ SM couplings from $M_t$ up to $M_P/\xi$ by using the three loop beta functions available in the literature \cite{RGE3}, and reproduced in a convenient form in the appendix of \cite{Buttazzo:2013uya}. For energies larger than $M_P/\xi$ the physical Higgs field decouples  and in this (relatively small) energy range we use the two loop RGE for $\lambda(\bar{\mu})$, $y_t(\bar{\mu})$, $g_3(\bar{\mu})$, $g_2(\bar{\mu})$ and $g_1(\bar{\mu})$  and  the one loop RGE for $\xi(\bar{\mu})$  in the chiral EW theory; these equations can be found for example by setting $s=0$ in the RGE given in  \cite{Allison:2013uaa}. The  use of the one loop RGE for $\xi$ (as opposed to the two loop ones for the SM parameters) will be justified in section \ref{numerical analysis}.

The RG-improvement used in this paper reaches a higher level of precision than the one in \cite{Bezrukov:2008ej}, where the running from the EW until the  $M_P/\xi$ scale was computed at  two loop  level  and then  one loop RGE were used at higher energies for all couplings.
\subsection{Threshold corrections}

We take the initial conditions at $\bar{\mu}=M_t$ for the $\overline{\rm MS}$ SM couplings from \cite{Buttazzo:2013uya},
which gives the most precise determination of the  threshold corrections for $\lambda$,  $g_3$ and $y_t$  available at the moment\footnote{Ref.  \cite{Buttazzo:2013uya} has improved on previous calculations (see Refs. \cite{Degrassi:2012ry} for the most recent ones).}:
\bea  \lambda(M_t)&=&0.12710+0.00206\left( \frac{M_h}{\rm GeV}-125.66 \right) \nonumber \\&&\hspace{-1cm} -0.00004 \left( \frac{M_t}{\rm GeV}-173.35\right)\pm
{0.00030}_{\rm th}, \nonumber \\ g_3(M_t)&=&1.1666 
+0.00314\frac{\alpha_3(M_Z)-0.1184}{0.0007} \nonumber \\ &&\hspace{-1cm} -0.00046 \left(\frac{M_t}{\rm GeV}-173.35 \right), \label{thresholds}\\ 
y_t(M_t) &=& 0.93697 +0.00550 \left( \frac{M_t}{\rm GeV}-173.35 \right) \nonumber  \\  &&\hspace{-1cm} 
 -0.00042\frac{\alpha_3(M_Z)-0.1184}{0.0007}\pm{0.00050}_{\rm th}.  \nonumber \eea 
We make use of these precise threshold corrections in our calculations.
The theoretical uncertainties on the quantities in (\ref{thresholds}) are much lower than those \cite{Kajantie:1995dw} used in previous determinations of the bound on $M_h$ from Higgs inflation (see a discussion on theoretical uncertainties in \cite{Bezrukov:2008ej}). 
 For the other couplings   $\alpha_2= g_2^2/(4\pi)$ and $\alpha_Y= 3g_1^2/(20\pi)$ we simply use the best fit  value from \cite{alpha}, which is precise enough for our purposes:
\be
\alpha_Y^{-1}(M_Z) =98.35\pm 0.013 ,\quad 
\alpha_2^{-1}(M_Z) = 29.587\pm 0.008 .
\ee
and extrapolate it to $M_t$ through the two loop SM RGE \cite{Buttazzo:2013uya} to obtain $g_1(M_t)\simeq 0.4631$ and $g_2(M_t)\simeq 0.6483$.
%

\section{Numerical studies}\label{numerical analysis}
In the numerical studies we use the following input parameters with corresponding uncertainties \cite{Buttazzo:2013uya}, \cite{Giardino:2013bma}:
\bea M_h&=&(125.66\pm 0.34) {\rm GeV} ,\nonumber \\  \quad M_t&=&(173.36 \pm 0.65\pm 0.3 ) {\rm GeV},  \label{inputs}\\ \alpha_3(M_Z)&=&0.1184\pm 0.0007. \nonumber \eea

Regarding the RGE, we connect the canonical SM   with the chiral EW theory regime by means of a  smooth function of the background field $\phi$, which is rapidly changing in the interval $[M_P/\xi,M_P/\sqrt{\xi}]$. The exact form of this function has a negligible impact on the numerics  and it can be taken to be the $s$ function first introduced by \cite{DeSimone:2008ei}.

As a first step in the numerical studies we determine $\xi$ taking into account quantum corrections. To do so we repeat the procedure summarized around Eqs. (\ref{normalization})-(\ref{inflation-end}), but with the classical potential replaced by the effective one: we substitute $U\rightarrow U_{\rm eff}$ in Eqs (\ref{normalization})-(\ref{inflation-end}) as well as in the definition of $\epsilon$ in Eq. (\ref{epsilon-def}). For numerical convenience we choose $\xi_t=\xi(M_t)$ and compute $\xi_{\rm inf}\equiv \xi(M_P/\sqrt{\xi_t})$ through the RGE. However, strictly speaking the running of $\xi$ from the EW up to the inflationary scale is not needed: the above mentioned procedure already gives this parameter for $\bar{\mu}\sim M_P/\sqrt{\xi_t}$. The running during the inflationary epoch spans a relatively small energy range and $\xi$ changes slowly as compared to the  relevant SM parameters; this justifies  the use of the one loop RGE for $\xi$. 

\begin{figure}[ht]
\includegraphics[scale=0.6]{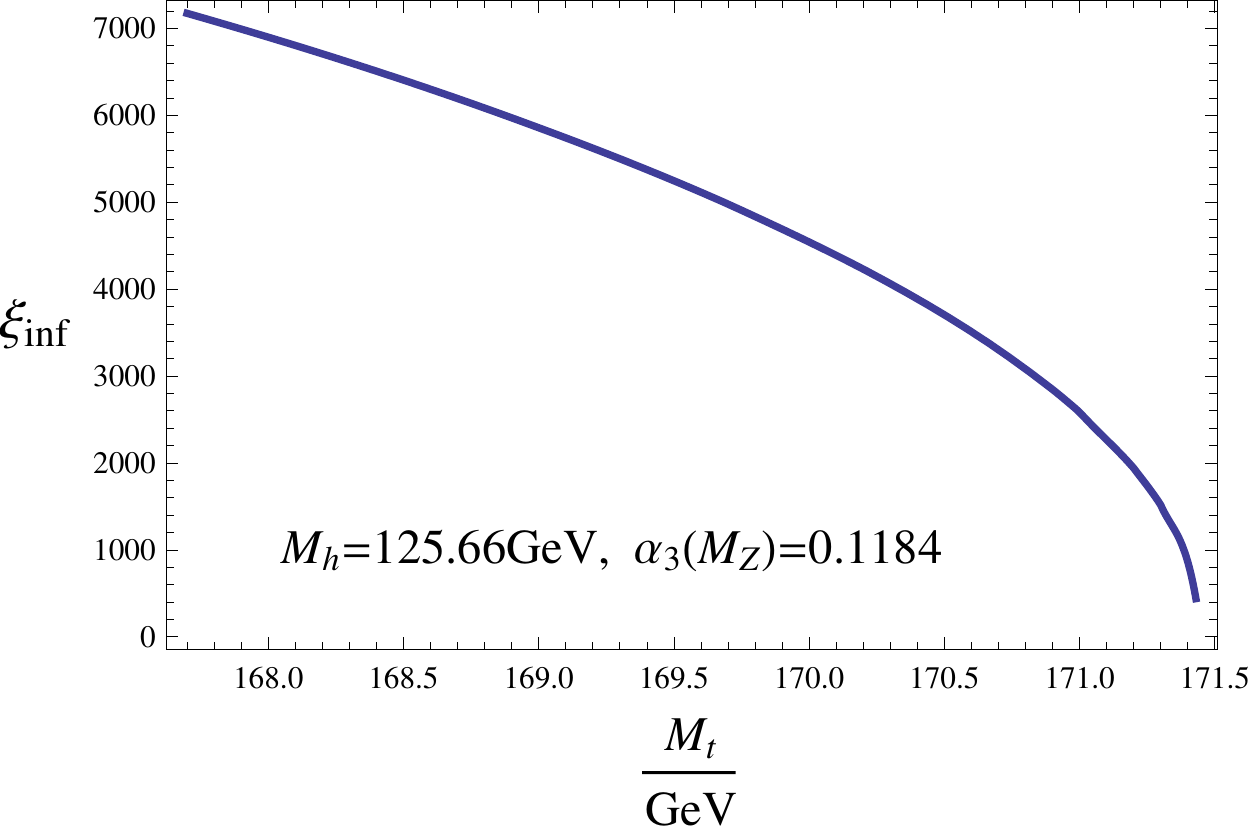}  
   \caption{{\small $\xi$ at the inflationary scale (defined in the text) versus the pole top mass $M_t$ setting to zero the theoretical uncertainties in Eqs. (\ref{thresholds}).}}
\label{xivsMt}
\end{figure}

We give $\xi_{\rm inf}$ as a function of $M_t$ in Fig. \ref{xivsMt} for some values of the input parameters. 
However, one should keep in mind that varying   $y_t(M_t)$ and $\lambda(M_t)$ within their theoretical uncertainties as well as $M_h$ and $\alpha_3(M_Z)$ within their errors (see Eq. (\ref{inputs})) visibly changes this plot. The changes produced by varying the WMAP normalization in (\ref{normalization}) by the 1$\sigma$ uncertainty reported in \cite{Hinshaw:2012aka} are instead much smaller.

\begin{figure}[ht]
\hspace{1cm} \includegraphics[scale=0.45]{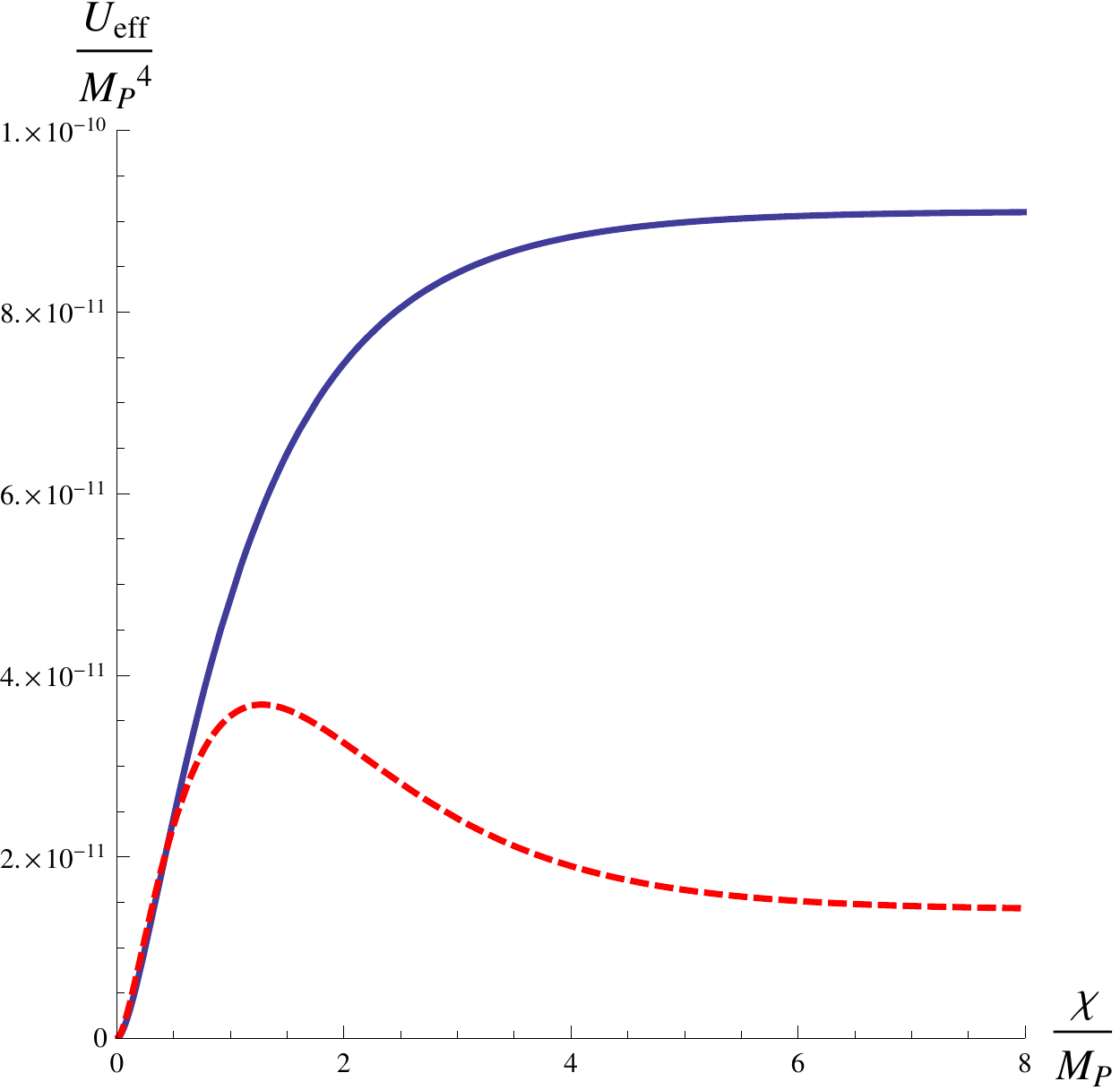}  
   \caption{{\small The effective Higgs potential setting to zero the theoretical uncertainties in Eqs. (\ref{thresholds}) and taking the central values in Eqs. (\ref{inputs}) except for the top pole mass $M_t$: we set $M_t=171.43$GeV in the solid line and $M_t=171.437$GeV in the dashed line. In the former case $\xi$ is fixed as described in the text, while in the latter case we set $\xi_t=300$. }}
\label{potential}
\end{figure}

Once $\xi$ is fixed, we can obtain our desired bound on $M_h$ (or $M_t$). As we mentioned in the introduction this bound can be obtained by requiring the slope of the Higgs potential to remain positive at energies around the inflationary scale. We illustrate this point in Fig. \ref{potential}, where we take $M_t$ to be the maximum  value to have inflation (fixing the other relevant parameters) or a bit larger; in the latter case the slope of the effective potential becomes negative because a bump develops at $\chi \sim M_P$. The effective potential in Fig. \ref{potential} is the NNLO one including the RG-improvement as described in section \ref{Quantum corrections}.

We find the following bound:
\bea \hspace{-0.7cm}\frac{M_h}{{\rm GeV}} > 129.46 +\frac{M_t - 173.36 {\rm GeV}}{0.50746 {\rm GeV}} \nonumber \\ \label{boundMh1}\hspace{0.5cm}   - 0.542 \frac{\alpha_3(M_Z) - 0.1184}{0.0007} \pm 0.23_{\rm th}.   \eea
Notice that this is a bit weaker than the one found in the second paper of \cite{Bezrukov:2008ej} (for prescription I): setting $M_t=171.2$ and $\alpha_3(M_Z)=0.1176$ we obtain $M_h/{\rm GeV}> 125.83  \pm 0.23_{\rm th}$ that is lower than $126.1$GeV. Here the main improvement with respect to the result in \cite{Bezrukov:2008ej} is the small theoretical uncertainty $0.23_{\rm th}$; this has to be compared with the value found in the previous paper, $2.2_{\rm th}$, which is above the error\footnote{The value 1.5 is obtained combining in quadrature the error on $M_t$ and $\alpha_3(M_Z)$.}, $1.5$, due to the  {\it current}  uncertainties on $M_t$ and $\alpha_3(M_Z)$. 

Combining in quadrature the experimental and theoretical uncertainties we obtain 
\be \frac{M_h}{{\rm GeV}} > 129.46 \pm 1.53.\label{boundMh2} \ee

Since the uncertainty on $M_h$ is already smaller than the one on $M_t$, it is useful to translate this bound into an upper bound on $M_t$:
\bea \frac{M_t}{\rm GeV} < 171.43 + 0.5075 \left(\frac{M_h}{\rm GeV} - 125.66\right) \nonumber \\ + 0.275\frac{\alpha_3(M_Z) - 0.1184}{0.0007}  \pm 0.117_{\rm th} \label{boundMt1} \eea
and combining in quadrature the experimental and theoretical uncertainties
\be \frac{M_t}{\rm GeV}< 171.43 \pm 0.35, \label{boundMt2} \ee
which is slightly weaker than the bound to have absolute stability of the EW vacuum in the pure SM \cite{Buttazzo:2013uya}, although the difference is well within $1\sigma$ uncertainty. The fact that this bound is weaker was expected because the prescription in (\ref{prescriptionI}) tells us that only the running up to $M_P/\sqrt{\xi_t}$ is relevant and there are cases in which the effective potential of the SM becomes smaller than its value at the EW vacuum only above $M_P/\sqrt{\xi_t}$. Regarding the proximity of (\ref{boundMt2}) to the above-mentioned bound in \cite{Buttazzo:2013uya}, we do not find any clear way to tell a priori that it should be so pronounced (within 1$\sigma$ uncertainty) and therefore we regard it as the result of explicit calculations. Therefore, with the present paper, we have shown  that even at the level of precision considered, the bound to have successful Higgs inflation  is essentially the same as the one to have stability of the EW vacuum in the pure SM.

\begin{figure}[ht]
\includegraphics[scale=0.6]{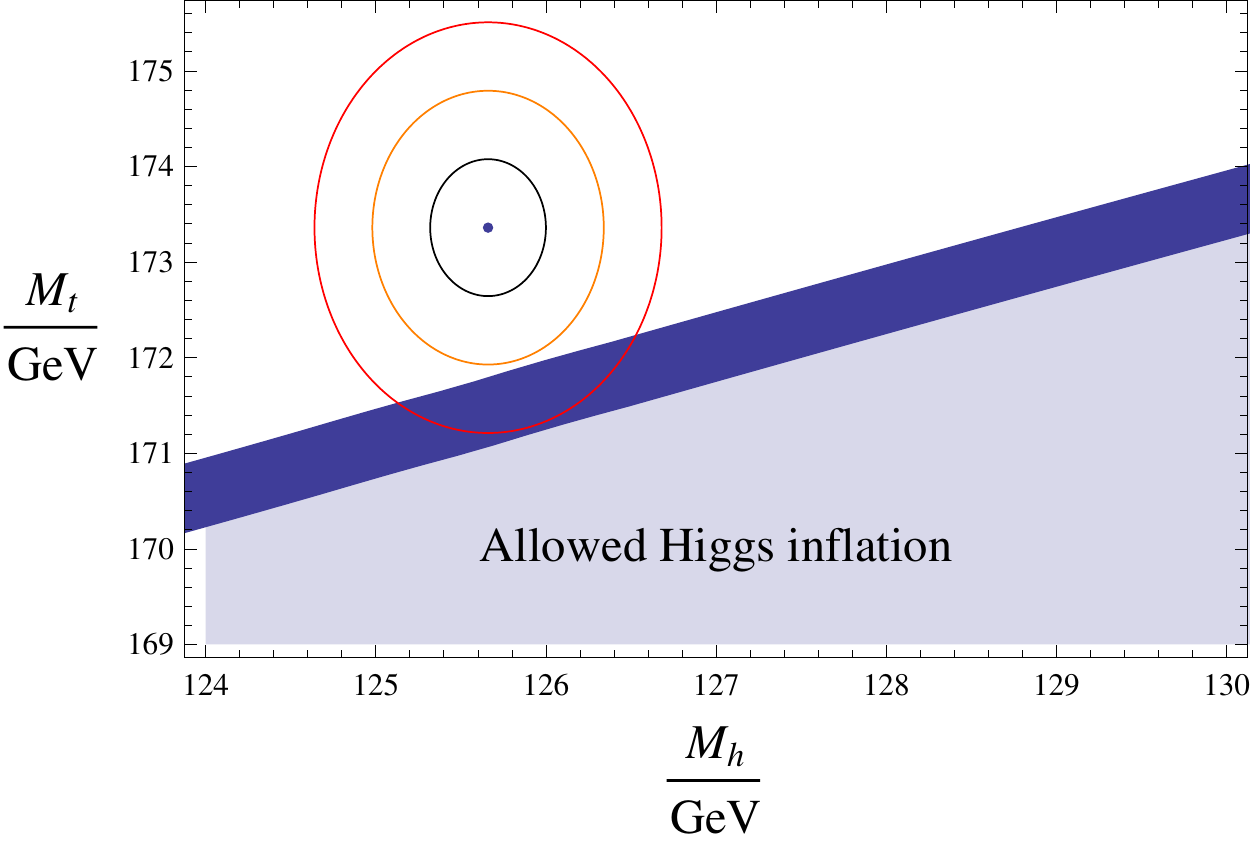}  
   \caption{{\small The upper bound on the top pole mass $M_t$ as a function of the top Higgs mass $M_h$. The width of the dark blue stripe is the 1$\sigma$ uncertainty, which is mainly due to the uncertainty on $\alpha_3(M_Z)$. Such width is basically the result of combining  in quadrature the uncertainties in Eq. (\ref{boundMt1}), except the one on $M_h$, which here is an independent variable. We also provide the experimental values of $M_h$ and $M_t$ with the ellipsis corresponding to the 1,2 and 3 $\sigma$ uncertainties.}}
\label{MhvsMt}
\end{figure}

In Fig. \ref{MhvsMt} we provide this bound as a function of $M_h$.   
One can see a tension with the experimental values of $M_h$ and $M_t$  but the overlap between the uncertainties on these masses and that of the bound itself is too big to exclude Higgs inflation. Using the precise threshold corrections in (\ref{thresholds}) we find that the uncertainty on the bound (the width of the blue line in Fig. \ref{MhvsMt}) is mainly due to the uncertainty on $\alpha_3(M_Z)$. This should be viewed as an improvement with respect to previous   determinations \cite{Bezrukov:2008ej}, because there the theoretical uncertainties on $\lambda(M_t)$ and $y_t(M_t)$ had an impact larger than the one we have here on $\alpha_3(M_Z)$. Indeed, if we were to use the theoretical uncertainty found in \cite{Bezrukov:2008ej} (see discussion below equation (\ref{boundMh1})) we would have a theoretical uncertainty about ten times bigger than the  one in Eq. (\ref{boundMt1}), which would result in a blue stripe about four times thicker than the one in Fig.  \ref{MhvsMt} and we would not see any tension.

Also, we find  that the uncertainty (see \cite{Hinshaw:2012aka}) on the WMAP normalization in (\ref{normalization}) has a negligible impact on these bounds.

We also computed the parameters $n_s$, $dn_s/d\ln k$ and $r$ making use of the full NNLO effective potential, that is by replacing $U\rightarrow U_{\rm eff}$ in Eqs. (\ref{epsilon-def}). We find values very close to previous (less precise) determinations \cite{Allison:2013uaa}, that were already perfectly compatible with recent Planck constraints \cite{Ade:2013uln}; for this reasons we do not display them  here.

Using prescription II for the quantization one just substitutes the value of $\bar{\mu}$ in (\ref{prescriptionI}) with $\bar{\mu}=\phi$ \cite{Bezrukov:2008ej}. As expected, this leads to a stronger bound: using the central values in (\ref{inputs}) and setting the theoretical uncertainties in (\ref{thresholds}) to zero we get $M_h>130.4$GeV  and $M_t<171.2$GeV: if this bound is not fulfilled we find that $n_s$ goes out of the allowed region provided by recent Planck results \cite{Ade:2013uln}, $n_s \gtrsim 0.94$.

\section{Conclusions}\label{conclusions}

In this paper we have derived the bound on $M_h$ (or equivalently $M_t$) in order for the Higgs field to be a viable inflaton, by including NNLO corrections (as discussed in the introduction). The bound on the Higgs mass is given in Eqs. (\ref{boundMh1}) and (\ref{boundMh2}), while the reformulation in terms of $M_t$ can be found in Eqs. (\ref{boundMt1}) and (\ref{boundMt2}).
We found a bound a bit weaker than previous determinations, Ref. \cite{Bezrukov:2008ej}. However,  the main improvement is not the central value of the bound, but its theoretical uncertainty, which has been reduced of one order of magnitude, and moved below the current uncertainties on $M_t$ (or $M_h$) and $\alpha_3(M_Z)$.  Moreover, the bound we found is  slightly weaker than the bound (obtained with the same level of precision, \cite{Buttazzo:2013uya}) coming from the requirement of absolute stability of the EW vacuum in the SM, although the difference is well within 1$\sigma$ uncertainty.  For this reason, it turned out that they can be essentially identified. 

We also provided a plot with such bound in the $M_h$-$M_t$ phase diagram (see Fig. \ref{MhvsMt}), where the width of the blue stripe represents the 1$\sigma$ uncertainty due to theoretical and experimental errors; since we made use of the currently most precise threshold corrections at the EW scale, Eqs. (\ref{thresholds}), this width is mainly due to the uncertainty on $\alpha_3(M_Z)$. The stripe is roughly 2-3$\sigma$ away from the present experimental values of $M_h$ and $M_t$. Therefore, while the Higgs inflation proposal  is not yet excluded, this reveals some tension with the  experiments. It would have not been possible to observe such tension if we had the same theoretical uncertainty as in the previous determination of \cite{Bezrukov:2008ej}, which we regard as the main reason why our computation is useful.

In passing we have also computed parameters of cosmological interest, such as $\xi$ (which is given as a function of $M_t$ in Fig. \ref{xivsMt}), $n_s$, $dn_s/d\ln k$ and $r$. We found that Higgs inflation  fulfills the observational constraints of the recent Planck release  \cite{Ade:2013uln}, even after full NNLO corrections are taken into account.

In order to make progress in understanding the viability of this proposal it is therefore crucial to reduce the uncertainties on the relevant parameters, in particular  $M_t$ and $\alpha_3(M_Z)$.

\vspace{1cm}

\noindent{\bf Note added:} after this paper was posted on the arXiv, Ref.  \cite{Allison:2013uaa} was updated and now it agrees with our expressions for $h$ and $g$.

\section*{Acknowledgments} We thank  Mikhail Shaposhnikov for valuable correspondence, Juan Garc\'ia Bellido and Juan Jos\'e Sanz-Cillero for useful discussions   and Elisabetta Majerotto for help in the computer codes. This work has been supported by the Spanish Ministry of Economy and Competitiveness under grant FPA2012-32828, 
Consolider-CPAN (CSD2007-00042), the grant  SEV-2012-0249 of the ``Centro de Excelencia Severo Ochoa'' Programme and the grant  HEPHACOS-S2009/ESP1473 from the C.A. de Madrid.









\end{document}